\documentclass[12pt]{article} 
\usepackage{amssymb}              
\newlength{\dinwidth}
\newlength{\dinmargin}
\newcommand{\AdS}{\mbox{{AdS}$_n\,$}}  
\newcommand{\RR}{{\bf R}}
\newcommand{\CC}{{\bf C}}
\newcommand{\NN}{{\bf N}}
\newcommand{\AdSG}{\mbox{SO$_0(2,n\!-\!1)$}}
\newcommand{\AdSGG}{\mbox{SO$(2,n\!-\!1)$}}
\newcommand{\CA}{{\cal A}}
\newcommand{\CH}{{\cal H}}
\newcommand{\CO}{{\cal O}}
\newcommand{\CW}{{\cal W}}
\newcommand{\rva}{{| 0 \rangle}}
\newcommand{\lva}{{\langle 0 |}}
\newcommand{\rphi}{{| \,\phi \rangle}}
\newcommand{\lphi}{{\langle \phi |}}
\begin{document}
\title{The Second Law of Thermodynamics, TCP, and Einstein Causality
in Anti--de Sitter Space--Time}
\author{Detlev Buchholz$^a$, \ 
Martin Florig$^b$ \ and \ Stephen J.\ Summers$^b$  \\[3mm] 
$^a$ \ Institut f\"ur Theoretische Physik,
Universit\"at G\"ottingen, \\ 
D--37073 G\"ottingen, Germany \\[3mm]
$^b$ \ Department of Mathematics, University of Florida, \\ 
Gainesville FL 32611, USA}
\date{September, 1999}
\maketitle
%
%
\begin{abstract} \noindent
If the vacuum is passive for uniformly accelerated observers in anti--de 
Sitter space--time ({\em i.e.} cannot be used by them to operate a 
{\em perpetuum mobile}), they will (a) register a 
universal value of the   
Hawking--Unruh temperature, (b) discover a TCP symmetry, and (c) find that 
observables in complementary wedge--shaped regions  
are commensurable (local) in the vacuum state. These results are 
model independent and hold in any theory which is compatible with 
some weak notion of space--time localization. 
\end{abstract}
\vspace*{4mm}
Quantum field theory in anti--de Sitter space--time (AdS) has recently 
received considerable attention \cite{Spires}. It seems therefore 
worthwhile to clarify in a model independent setting 
the universal properties of such theories  
which are implied by generally accepted and physically meaningful  
constraints. 

We report here on the results of an  
investigation of this question \cite{BuFlSu} which 
brought to light the fact that stability properties of the 
vacuum (the impossibility of operating a {\em perpetuum mobile})
are at the root of (a) the Hawking--Unruh effect in AdS,  
(b) a TCP--like symmetry in AdS, and (c) local commutativity 
properties of the observables. As will be explained, 
these results hold in any quantum theory in AdS which allows 
one in principle to specify the spacetime localization 
properties of observables. 

We consider here AdS of any dimension $n \geq 3$. It  can conveniently be
described in terms of Cartesian coordinates in the ambient space
$\RR^{n+1}$  as the quadric
$$
\AdS = \{x\in\RR^{n+1} : 
x^2 \equiv x_0^2 - x_1^2 - \dots - x_{n-1}^2 + x_n^2
= R^2 \}
$$ 
with metric $g = \mbox{diag}(1,-1,\dots,-1,1)$ in diagonal form. 
Its isometry group is O$(2,n-1)$ whose identity component will be  
denoted by \AdSG.

Any quantum theory in AdS has to determine the following  
structures, which are indispensible for its physical interpretation: 

\noindent (i) \ A Hilbert space $\CH$, describing the physical states,
and a continuous unitary representation $U$ of 
the symmetry group \AdSG, acting on $\CH$.\footnote{It is sufficient here to 
consider the subspace of ``bosonic'' states, so we do not 
need to proceed to the covering group of the space--time 
symmetry group. \label{1}}

\noindent (ii) \ A set of operators which
describe the observables of the theory. Since these operators can 
be added and multiplied, they generate an algebra $\CA$ which  
will be called the algebra of observables.\footnote{We work in the 
framework of  von Neumann algebras without dwelling on
this point any further. For more information on the
mathematical background, cf. \cite{Ha}. \label{2} } 

It is a fundamental feature of quantum field theory and its modern 
ramifications that the spacetime localization properties of 
observables enter into the formalism in an essential way.
One can tell whether an observable is localized about  
a spacetime point, a string, loop, disk {\it etc.} Thus, given any  
region $\CO \subset \AdS$, one can identify   
those observables which are localized within that region
and define a corresponding subalgebra $\CA (\CO) \subset \CA$ 
which is generated by them. Since there exist no fewer observables 
if the region becomes larger, one obviously has
$$
\CA (\CO_1) \subset \CA (\CO_2) \quad \mbox{if}  \quad \CO_1 \subset
\CO_2.
$$

The localization of observables has 
to be compatible with the action of the 
spacetime symmetry group. If an observable $A$ 
is localized in the region $\CO$, say, its image 
$U(\lambda) A U(\lambda)^{-1}$ under the action of 
$\lambda \in \AdSG$  
should be localized in the transformed region $\lambda \CO$. 
The transformation properties of the 
observables under the action of the 
unitaries $U(\lambda)$ may be quite complicated, but 
fortunately this detailed information is not needed here. All that matters 
are the geometrical aspects of the transformations, which can be 
summarized as follows.

\noindent (iii) For each region $\CO\subset \AdS$ 
and $\lambda \in \AdSG$
one has the equality (in the set--theoretic sense) 
$$
U(\lambda)  \CA (\CO) U(\lambda)^{-1} = \CA (\lambda \CO).  
$$

We emphasize that we do not postulate here  
from the outset local commutation relations of 
the observables. For, in contrast to 
the case of globally hyperbolic space--times,
the principle of Einstein causality does 
not provide any clues 
as to which observables in AdS should commute. 
Instead, we will derive such commutation relations from 
stability properties of the vacuum. 
Since this stability aspect sheds also new 
light on the phenomenon of the Hawking--Unruh 
temperature, we explain this point in 
somewhat more detail.

We begin by describing the observers in AdS which are of 
interest here. Let $x_O \in \AdS$ be any point and let
$\lambda (t)$, $t \in \RR$, be any one--parameter subgroup of
$\AdSG$ such that $t \rightarrow \lambda (t) x_O$ is an 
orthochronous curve. 
We interpret this curve as the worldline of some observer
(assumed to be male for concreteness). Depending on the 
initial data chosen, this observer will be geodesic or
experience some constant acceleration. Points in the 
neighborhood of $x_O$ will in general also give rise 
to orthochronous curves under the action of the 
chosen subgroup of $\AdSG$, and we denote by $\CW$ the 
connected neighborhood of $x_O$ in \AdS consisting of all
such curves. The region $\CW$, which can be all of AdS
or a subset of it, is that part of the space--time in which 
the observer can perform measurements. The corresponding
observables are described by the self--adjoint elements of 
$\CA(\CW)$ and the corresponding dynamics is given by 
$e^{\, itM} \equiv U(\lambda (t)) $ with generator (``Hamiltonian'')
$M$.\footnote{We choose a fixed parametrization of 
the pertinent subgroups of \AdSG. The proper time of the 
observer is obtained by rescaling $t$ with
$((\dot{\lambda}(0) \, x_O)^2){}^{1/2}$. \label{4}}

Let us discuss next what this observer can say about the 
properties of the vacuum, described by the 
unit vector $\rva \in \CH$. First, he can 
check whether the vacuum is  
suitable to operate a {\it perpetuum mobile\/} of 
the second kind. To this end he 
would perform all kinds of operations in $\CW$ which 
can be described by perturbations of the generator $M$ in 
that region. Time--dependent perturbation theory
then implies that the state of the vacuum changes 
under the influence of such perturbations and is 
described by a vector $V \rva$  for some unitary operator 
$V \in \CA (\CW)$ after the perturbation has been turned off.
Since this challenge of the Second Law of Thermodynamics ought to fail, the 
observer should find that, no matter what he does, the 
energy of the final state is no less than that of
the initial vacuum, {\em i.e.} he cannot extract energy from the vacuum. 
In formula form, 
$$
\lva \, V^* M V \, \rva \geq \lva \, M \, \rva 
$$
for all {\it unitary\/} operators $V \in \CA (\CW)$. This 
property of a state --- not being able to perform work in 
a cyclic process --- is called passivity \cite{PuWo}. 
It is a mathematical expression of the Second Law 
of Thermodynamics .

The observer can also test with the help of time 
averages of his observables (yielding order parameters) 
whether the vacuum is a mixture of different phases. Since the 
vacuum is the most elementary system, the answer should be 
negative, {\em i.e.} all order parameters should have sharp values. 
This is the content of the relation (mixing property)
$$
\lim_{T \rightarrow \infty}
\frac{\mbox{\footnotesize $1$}}{\mbox{\footnotesize $T$}}
\int_0^T \! dt \, \big( \lva A(t) B \rva  -  
\lva A(t) \rva \, \lva  B  \rva \big) = 0 
$$
for $A,B \in \CA(\CW)$, where  
$A(t) \equiv e^{\, itM } A e^{\, -itM }$.
These basic features of the vacuum can be summarized  
as follows.\\[1mm]
(iv) The vacuum $\rva$ is, for all geodesic and all uniformly 
accelerated observers, a passive and mixing state.

Let us now turn to the analysis of the implications 
of these assumptions. 
Here we profit from a deep result of Pusz and Woronowicz 
for arbitrary quantum dynamical systems~\cite{PuWo}. In the present 
context this result says that the vacuum vector $\rva$ is, 
as a consequence of its passivity and mixing properties,  
invariant under the dynamics of the above observer, {\em i.e.}\ 
$M \rva = 0$, and it is either a ground state for $M$, 
or satisfies, for some {\it a priori\/} unknown $\beta \geq 0$, 
the Kubo--Martin--Schwinger (KMS) condition, which can 
be presented in the form\footnote{Since $e^{\, - \beta M}$ is in general an 
unbounded operator, this relation is to be understood in the sense of 
quadratic forms. \label{5}} 
$$
\lva B e^{\, -\beta M} A^* \rva = \lva A^* B \rva ,
$$
for all $A,B \in \CA(\CW)$. In the latter case the 
respective observer would interpret the vacuum 
as a thermal equilibrium state at temperature $\beta^{-1}$. 

It is convenient here to proceed to another formulation of the 
above result. Making use of the fact that  
$\lva A^* B \rva = \overline{\lva B^* A \rva}$ and that 
the set of vectors $\CA ( \CW) \, \rva$ is dense in $\CH$
\cite{BuFlSu}, the above relation allows one to define an anti--unitary
operator $J$ on $\CH$ by setting\footnote{The operator $J$ is called 
modular conjugation in Tomita--Takesaki theory.  
Some of our explicit computations could be abbreviated by 
using general results from this theory. \label{6}}
$$
J A \rva \equiv e^{\, -(\beta/2) M } A^* \rva \ \ 
\mbox{for} \ \ A \in \CA(\CW).
$$

Our main task will be to determine 
the specific properties of this operator.  
In order to simplify the necessary computations,  
we consider the particular choice of region
$$
\CW = \{ x \in \AdS : x_1 > |x_0|\, , x_n > 0 \} 
$$
on which the one--parameter subgroup of 
boosts $\lambda_{0 1} (t)$, $t \in \RR$, 
in the $0$--$1$--plane acts in an  
orthochronous manner. For any $x_O \in \CW$, the curve 
$t \rightarrow \lambda_{0 1} (t) x_O$ is the worldline of some 
observer, as described above. 

We shall determine the temperature felt by this
observer. Since the generator $M_{0 1}$ of 
his dynamics is transfomed into $- M_{0 1}$
by the adjoint action of the rotation 
$e^{\, i\pi M_{1 2}}$ in the $1$--$2$--plane, 
it cannot be a positive operator. Hence $\rva$ is not a 
ground state for this observer; it must therefore 
satisfy the KMS--condition for some $\beta \geq 0$. So what is the 
value of $\beta$ ? 

To answer this question, we apply the methods in 
\cite{BoBu}. We pick a region $\CO$ such that  
$\lambda_{0 2} (s) \, \CO \subset \CW$
for the boosts $\lambda_{0 2} (s)$ in the 
$0$--$2$--plane with sufficiently small parameter $s$. 
By the group law in \AdSG \, we have   
$$ e^{\, itM_{0 1}} e^{\, isM_{0 2}} = 
e^{\, is(\cosh(t)M_{0 2} + \sinh(t) M_{1 2})}  e^{\, itM_{0 1}}.  
$$
Taking into account that the boost $e^{\, isM_{0 2}}$ is the dynamics 
of some other observer, we infer from (iv)  
that $\rva$ is invariant under its action.  
Thus we get for any vector $\rphi \in \CH$ and 
operator $A^* \in \CA(\CO)$ 
\begin{eqnarray*}
\lefteqn{ \lphi \, e^{\, itM_{0 1}} 
\, e^{\, isM_{0 2}} A^* e^{\, -isM_{0 2}}  \rva}  
\\ &  & =  \lphi \, 
e^{\, is(\cosh(t)M_{0 2} + \sinh(t) M_{1 2})}\,  e^{\, itM_{0 1}} 
A^* \rva. 
\end{eqnarray*}

At this point we are in the position to combine the KMS--condition with  
a basic result in the representation theory
of Lie groups, due to Nelson \cite{Ne}. 
In the present situation this result can  
be stated as follows: There exists a dense set of vectors 
$\rphi \in \CH$  such that the corresponding vector functions
$$
u,v \rightarrow e^{\, i (u M_{0 2} + v M_{1 2})} \, \rphi 
$$ 
are analytic in a fixed neighborhood 
of the origin of $\CC$. 

Returning to the analysis of the above equality,
we note that 
according to condition (iii) the operator 
$e^{\, isM_{0 2}} A^* e^{\, -isM_{0 2}}$ 
appearing in the matrix element on the left hand side is an element of 
$\CA(\lambda_{0 2} (s)\CO) \subset \CA(\CW)$. So, in view of 
the KMS--property, we can continue this term analytically in $t$ to 
$i\beta/2$. In the expression on the right hand side 
there appears a product of two unitary operators 
which both depend on $t$.
If we choose for $\rphi$ a Nelson vector, as described above, the 
first operator, standing next to $\lphi$, can be analytically 
continued in $t$ to $i\beta/2$, provided $s$ is sufficiently 
small. The second operator, standing next to $ A^* \rva$, 
can likewise be continued to $i\beta/2$
by the KMS--property. So we arrive at   
\begin{eqnarray*}
\lefteqn{ \lphi \, e^{\, -(\beta/2)M_{0 1}} 
\, e^{\, isM_{0 2}} A^* e^{\, -isM_{0 2}}  \rva}  
\\ &  & =  \lphi \, 
e^{\, is(\cos(\beta/2) M_{0 2} + i \sin(\beta/2) M_{1 2})}  
e^{\, -(\beta/2)M_{0 1}} A^* \rva. 
\end{eqnarray*}
Bearing in mind the definition of 
$J$ and the invariance of the vacuum under 
the boosts $e^{\, -isM_{0 2}}$, we can bring this 
relation into the more transparent form
$$
\lphi \, J  
e^{\, isM_{0 2}} A \rva  =  \lphi \, 
e^{\, is(\cos(\beta/2) M_{0 2} + i \sin(\beta/2) M_{1 2})} J  A \rva.
$$ 
As this holds for the dense sets of vectors 
$\rphi$ and $A \rva$, respectively, we conclude that 
$$
J  e^{\, isM_{0 2}}   =   
e^{\, is(\cos(\beta/2) M_{0 2} + i \sin(\beta/2) M_{1 2})} J.
$$
The operator on the left hand side is anti--unitary, so the 
same must be true for the operator on the right hand side. This
is only possible if $\beta$ is an {\it integer\/} multiple of $2 \pi$,
for otherwise the operator appearing in the 
exponential function is not skew-adjoint. By a more 
refined functional analytic argument one can restrict $\beta$ 
even further and show that its only possible value is 
$\beta = 2 \pi$ \cite{BuFlSu}. 

Proceeding to the proper 
time scale of the observer$^{\ref{4}}$,  
we conclude that he is exposed to the  
Hawking--Unruh temperature 
$(1/2\pi)((\dot{\lambda}_{0 1}(0)\, x_O)^2){}^{-1/2}$, 
in accordance with the value found in  
model computations \cite{DeLe} and also by more general 
considerations \cite{Ja}. 
By similar arguments one can compute the temperature
felt by other observers, such as   
geodesic ones, for whom it is zero \cite{BuFlSu}.
Thus we have established the following general fact: \\[2mm]
{\it Each uniformly accelerated observer 
testing the vacuum state in AdS finds
a universal value of the Hawking--Unruh temperature which depends only
on his particular orbit. For geodesic observers  the 
vacuum is a ground state.} \\[2mm]
This result is entirely a 
consequence of the passivity of the vacuum. Hence 
this state is the only one which is passive for all observers. 

Having computed the value of $\beta$, let us return now to the 
analysis of $J$. Plugging $\beta = 2 \pi$ into the above equality
for $J$, we see that for small $s$
$$
J  e^{\, isM_{0 2}}   =   
e^{\, - is  M_{0 2}} J.
$$
This relation can be extended to 
arbitrary $s$ by iteration, if one 
decomposes $e^{\, isM_{0 2}}$ into an $m$--fold product
$(e^{\, i(s/m)M_{0 2}})^m$ for sufficiently large $m$.

In a similar manner one 
can determine the intertwining properties of $J$
with other one--parameter subgroups of
\AdSG, and thereby with all unitaries $U(\lambda)$. 
We skip these computations and only state the result 
\cite{BuFlSu}: 
$$J  U(\lambda)   =   U(\theta \lambda \theta)  J, $$
where $\theta$ is the reflection which changes the sign of the 
$0$--$1$--coordinates of the points in \AdS. This result also 
allows us to show explicitly$^{\ref{6}}$ 
that $J^2 = 1$: For the intertwining 
relation and the anti--unitarity of $J$ imply
$$ 
J e^{\, - (\beta/2) M_{0 1}} = e^{\, (\beta/2) M_{0 1}} J,
$$
hence we have, in view of the defining equation for $J$, 
\begin{eqnarray*}
J^2 A \rva & = & J e^{ \, - (\beta/2) M_{0 1}} A^* \rva =
e^{\, (\beta/2) M_{0 1}}  J A^* \rva \\ & = &
e^{\, (\beta/2) M_{0 1}} e^{\, - (\beta/2) M_{0 1}} A^{**} \rva =
A \rva ,
\end{eqnarray*}
for all $A \in \CA(\CW)$. Summing up, we have established the 
following analogue of the TCP--theorem: \\[2mm]
{\it The representation $U$ of \AdSG extends to a 
representation of \AdSGG in which the reflection $\theta$
is implemented by the anti--unitary operator $J$.} \\[2mm]
This result is a purely group--theoretic statement
which does not yet say anything about the adjoint action
of $J$ on the observables. In order to gain insight 
into the nature of this latter action, we consider the observables 
which are localized in the region
$$
\CW^{\prime} \equiv 
\{x\in\AdS : - x_1 > | x_0 |, x_n > 0\}.
$$
Since the regions $\CW$ and $\CW^{\prime}$ are obtained by
intersecting  opposite wedge--shaped regions in the ambient 
space with \AdS, we call them {opposite wedges}. 

By a rotation in the $1$--$2$--plane about the angle $\pi$,
the region $\CW$ is transformed into $\CW^{\prime}$, and 
{\it vice versa}. Thus, if $A \in \CA(\CW)$ one has
$A^{\prime} \equiv e^{\, i \pi M_{1 2}} A 
e^{\, -i \pi M_{1 2}}\in \CA(\CW^{\prime})$, cf.\ (iii). 
By the intertwining relation for $J$, we have 
$J e^{\, i \pi M_{1 2}} = e^{\, -i \pi M_{1 2}} J = 
e^{\, i \pi M_{1 2}} J$, where we used in the second 
equality the fact$^{\ref{1}}$ that 
$e^{\, i 2 \pi M_{1 2}}  =1$. 
Finally, the group structure implies that 
$e^{\, i \pi M_{1 2}} e^{\, - (\beta/2) M_{0 1}} = 
e^{\, (\beta/2) M_{0 1}} e^{\, i \pi M_{1 2}}$. Combing these  
facts, we can compute  
\begin{eqnarray*}
J A^{\prime} \rva & = & J e^{\, i \pi M_{1 2}} A \rva 
= e^{\, i \pi M_{1 2}} J A \rva  \\ & = &
e^{\, i \pi M_{1 2}} e^{\, - (\beta/2) M_{0 1}} A^* \rva =
e^{\, (\beta/2) M_{0 1}} e^{\, i \pi M_{1 2}} A^* \rva  \\ & = &
e^{\, (\beta/2) M_{0 1}} A^{\prime *} \rva.
\end{eqnarray*}
Picking any $A^{\prime} \in \CA(\CW^{\prime})$, 
$B \in \CA(\CW)$, we obtain with the help of this result 
\begin{eqnarray*}
\lefteqn{\lva A^{\prime *} B \rva = \overline{\lva B^{*} A^{\prime}\rva}
= \lva B^{*} J^* J A^{\prime}\rva} \\ & & = 
\lva B^{* *} e^{\, -(\beta/2) M_{0 1}} 
e^{\, (\beta/2) M_{0 1}} A^{\prime *} \rva
= \lva B A^{\prime *} \rva.
\end{eqnarray*} 
Thus we find that the elements of $\CA(\CW)$ and 
$\CA(\CW^{\prime})$ commute in the vacuum state. 
It is apparent that the same statement holds also 
for the algebras corresponding to all transformed opposite 
wedges $\lambda \CW$ and $\lambda \CW^{\prime}$. This 
commutativity, in the vacuum state,  of observables which are 
localized in certain complementary regions is sometimes called 
{\it weak locality}. So we can state: \\[2mm]
{\it Observables which are localized in opposite wedges of AdS
are weakly local with respect to each other.} \\[2mm]
This establishes the asserted general relation between passivity 
properties of the vacuum and commensurability properties of 
observables. In theories 
where the basic fields satisfy c--number commutation relations, 
it follows from this result that observables 
localized in opposite wedges commute also in the 
usual (operator) sense, in accord with the structure found in 
concrete models \cite{Fr}.  

Before we turn to the discussion of our results, let us 
mention another property of $J$ which holds in all theories complying with 
the following stronger locality condition:\\[1pt]
(v) Observables which are localized in opposite wedges
of AdS commute with each other. \\[1mm]
This situation prevails in the models of current interest \cite{Rehren}. 
It implies that the operator $J$ implements 
the geometric action of the reflection $\theta$ also on the observables,
$$ 
J \CA (\CW) J^* = \CA (\theta \CW).
$$
As a matter of fact, the validity of this relation is equivalent to the  
strong locality condition. The proof of this result relies 
on methods of Tomita--Takesaki theory and will be given elsewhere
\cite{BuFlSu}. There it will also be shown that mild constraints on 
the mulitplicities of the eigenvalues $2 \pi \NN_0$ of the generator $M_{0 n}$ 
(existence of the partition function $\mbox{Tr} \, e^{\, - \beta
M_{0 n}}$) imply that 
observables in opposite wedges are statistically (causally)
independent \cite{BuWi}. 

We have thus established in this letter a tight relation between stability
properties of the vacuum, a TCP theorem, 
and commensurabilty properties of observables in 
AdS. Such relations exist also in other space--times, but 
they are of particular interest in the case of AdS. For they 
reveal that 
observables in opposite wedges $\CW$ and $\CW^{\prime}$ necessarily
commute with each other, either weakly or strongly. 

In the case of proper AdS, this result
appears to be very strange at first sight, since any point in $\CW$ can be
connected with any other point in $\CW^{\prime}$ by some causal 
curve. Hence measurements at these points should affect each other 
and not be commensurable (let alone be statistically independent). 

After a  moment's reflection, however, one sees that 
the regions $\CW$ and $\CW^{\prime}$, although causally connected in the 
usual sense, cannot be connected by causal geodesics. 
So the following picture emerges, which explains the unexpected 
commensurability properties of the observables: the perturbations 
which are produced by observers in the region $\CW$ move, after having 
left the region, along geodesics and therefore do not reach $\CW^{\prime}$.
In noninteracting theories this picture is reasonable, and free field
models show that it is also consistent. But it is in conflict
with the patterns of interaction processes which would inevitably  
lead to incommensurablilites of observables in opposite wedge regions. 
This clash between geometry and interaction can 
be evaded if one proceeds from proper AdS to its covering space,
where opposite wedges are causally disjoint. \\[6mm]
{\Large \bf Acknowledgement} \\[4mm]
The authors profitted from a  
correspondence with Jacques Bros. One of them (DB) 
would like to thank the University of Florida for hospitality and
financial support and the Deutsche Forschungsgemeinschaft DFG
for a travel grant.


\begin{thebibliography}{99}
\bibitem{Spires} We refrain from giving references and refer the 
interested reader to the HEP database, where 
a comprehensive list of articles on this topic 
may be retrieved.
\end{thebibliography}
\end{document}